# Asymptotic analysis of a line source diffraction by a perfectly conducting half-plane in a bi-isotropic medium.

**W. Hussain***

*Department of Mathematics, School of Arts and Sciences, Lahore University of Management Sciences (LUMS), Opposite Sector'U', D.H.A., Lahore Cantt. 54792, Pakistan.*

* *E-mail address*: wasiq@lums.edu.pk (W.Hussain)
Tel: 92-42-5722670-79(10 lines); fax: 92-42-5722591

## Abstract

This paper is concerned with the diffraction of an electromagnetic wave by a perfectly conducting half-plane in a homogeneous bi-isotropic medium (*asymptotically*). Similar analysis in a source-free field is done in Asghar,S. and Lakhtakia, A., (1994), Plane-wave diffraction by a perfectly conducting half-plane in a homogeneous bi-isotropic medium. Int. J Appl. Electromagnetics in materials,5,(1994), 181-188.

*In this paper attention is focused on the wave coming from a **line source**.*

The objective is to study the scattering of an electromagnetic wave from the boundary of a half-plane and thereby to provide a theoretical framework for the *line source diffraction asymptotically*. In far field approximation it is shown that an incident wave coming from a line source behaves like a plane wave. The scattered field is calculated by using the Fourier transform and the Wiener-Hopf techniques. The scattered field in the far zone is determined by using contour integration.

## 1. Introduction

The study of Beltrami fields goes back to the 19$^{th}$ century (Ref. [4]). The details of Beltrami fields in chiral media (reciprocal bi-isotropic media) are available in Ref. [8]. Analysis of electromagnetic plane-wave diffraction by a metallic strip is established in Ref. [5]. The study of Beltrami fields plays an important role in the description of time-harmonic electromagnetic fields in the bi-isotropic media (Ref. [20]). The main difference between the Beltrami-Maxwell formalism and the conventional electromagnetic theory of Maxwell lies in the nature of the fields. The fields appearing in the time-dependent Maxwell's equations are real-valued, whereas Beltrami fields are complex-valued, while a close relation between the two can be established, see Ref. [10].

Texts are also available giving a detailed study of Beltrami fields and chiral media, for example, Ref. [7]. The availability of Beltrami-Maxwell equations for general material continua opens the door to research problems in electromagnetics. The simplest continuum that can exist is vacuum and solutions therein are possible for Beltrami fields, both in source–free problems as well as in the radiation problems involving confined sources.

Asghar and Lakhtakia in [1] analyzed the *source-free* plane-wave diffraction by a perfectly conducting half-plane in a homogeneous bi-isotropic medium. They reduced the vector diffraction problem to the scattering of a single scalar field and calculated the left-handed Beltrami field since the right-handed Beltrami field could be found in a similar way.

In this paper the analysis done in [1] is extended asymptotically to the problem of a ***Line Source.*** *Asymptotic techniques are sometimes very useful in tackling problems of complicated nature.* See for example References ([6], [18]).

The required notations and equations are summarized in Section 2 along with a brief review of [1] in the sub-section 2.1 In Section 3 the problem in [1] is extended to the line source diffraction *asymptotically*. It is shown that an incident wave coming from a line source behaves like a plane wave when the line source is shifted to infinity. In Section 4 the Wiener-Hopf equations for the scattered field are given and their solution is discussed in the sub-section 4.1. The field in the far zone is evaluated in the sub-section 4.2 and finally the conclusions are given in Section 5.

## 2. Basic equations

Let us assume that the whole space is occupied by a homogeneous bi-isotropic medium except for a perfectly conducting half-plane $z = 0, x \geq 0$ . In the Federov representation (Ref. [7]), the bi-isotropic medium is characterized by the following equations

$$\underline{D} = e\underline{E} + ea\underline{\nabla} \times \underline{E}\,, \qquad (2.1)$$

$$\underline{B} = m\underline{H} + mb\underline{\nabla} \times \underline{H}\,, \qquad (2.2)$$

where $e$ and μ are the permittivity and the permeability scalars respectively, while α and β are the bi-isotropy scalars. $\underline{D}$ is the electric displacement, $\underline{H}$ is the magnetic intensity, $\underline{B}$ is the magnetic flux density, and $\underline{E}$ is the electric intensity.

Let us assume the time-dependence of Beltrami fields to be of the form $\exp(-iwt)$, where $w$ is the angular frequency. The source-free Maxwell curl postulates in the bi-isotropic medium can be set up as

$$\underline{\nabla} \times \underline{Q}_1 = g_1 \underline{Q}_1\,, \qquad (2.3)$$

$$\underline{\nabla} \times \underline{Q}_2 = -g_2 \underline{Q}_2\,, \qquad (2.4)$$

where Beltrami fields are given as in Ref. [9]

$$\underline{Q}_1 = \left( \frac{h_1}{h_1 + h_2} \right) \left[ \underline{E} + \mathrm{i} h_2 \underline{H} \right], \tag{2.5}$$

and $$\underline{Q}_2 = \left( \frac{h_1}{h_1 + h_2} \right) \left[ \underline{H} + \mathrm{i} \frac{\underline{E}}{h_1} \right], \tag{2.6}$$

in terms of the electric field $\underline{E}$ and the magnetic field $\underline{H}$.

The two wave numbers appearing in (2.3) and (2.4) are given by

$$g_1 = \frac{k}{(1 - k^2 a b)} \left\{ \sqrt{1 + \frac{k^2 (a-b)^2}{4}} + \frac{k(a+b)}{2} \right\}, \tag{2.7}$$

$$g_2 = \frac{k}{(1 - k^2 a b)} \left\{ \sqrt{1 + \frac{k^2 (a-b)^2}{4}} - \frac{k(a+b)}{2} \right\}, \tag{2.8}$$

and two impedances in (2.5) and (2.6) are given by

$$h_1 = \frac{h}{\left\{ \sqrt{1 + \frac{k^2 (a-b)^2}{4}} + \frac{k(a-b)}{2} \right\}}, \tag{2.9}$$

and

$$h_2 = \frac{h}{\left\{ \sqrt{1 + \frac{k^2 (a-b)^2}{4}} + \frac{k(a-b)}{2} \right\}}, \tag{2.10}$$

where $\kappa = \sqrt{em}$ and $\eta = \sqrt{\frac{m}{e}}$.

Here $\underline{Q}_1$ is the left-handed Beltrami field and $\underline{Q}_2$ is the right-handed Beltrami field.

### 2.1 Review of source-free diffraction problem in [1].

In this sub-section a brief review of the diffraction problem in [1] is given. In [1]

*scattering along the **y**-direction is considered.*

Writing the Beltrami field $\underline{Q}_1$ as [21]

$$\underline{Q}_1 = Q_{1x}\underline{i} + Q_{1y}\underline{j} + Q_{1z}\underline{k}. \tag{2.11}$$

and $Q_{1x}\underline{i} + Q_{1z}\underline{k} = \underline{Q}_{1t}$, we have

$$\underline{Q}_1 = \underline{Q}_{1t} + \underline{j}Q_{1y}. \tag{2.12}$$

Writing $\underline{j} = \hat{y}$ i.e. unit vector along the $y$-axis we have

$$\underline{Q}_1 = \underline{Q}_{1t} + \hat{y}Q_{1y}. \tag{2.13}$$

Similarly we can write down $\underline{Q}_2$ as

$$\underline{Q}_2 = \underline{Q}_{2t} + \hat{y}\underline{Q}_{2y}. \tag{2.13a}$$

Clearly the fields $\underline{Q}_{1t}$ and $\underline{Q}_{2t}$ lie in the $XZ$- plane and $\hat{y}$ is a unit vector along the $y$-axis such that $\hat{y}.\underline{Q}_{1t} = 0$ and $\hat{y}.\underline{Q}_{2t} = 0$.

The equation (2.3) can be written as

$$\begin{vmatrix} \underline{i} & \underline{j} & \underline{k} \\ \frac{\partial}{\partial x} & \frac{\partial}{\partial y} & \frac{\partial}{\partial z} \\ Q_{1x} & Q_{1y} & Q_{1z} \end{vmatrix} = g_1\left(Q_{1x}\underline{i} + Q_{1y}\underline{j} + Q_{1z}\underline{k}\right).$$

Now assuming all the field vectors having an implicit $\exp(ik_y y)$ dependence on the variable $y$ and comparing $x$ and $z$ components on both sides of the above equation we have

$$Q_{1x} = \frac{1}{k_{1xz}^2}\left[ik_y \frac{\partial Q_{1y}}{\partial x} - \gamma_1 \frac{\partial Q_{1y}}{\partial z}\right] \tag{2.14}$$

and

$$Q_{1z} = \frac{1}{k_{1xz}^2}\left[ik_y \frac{\partial Q_{1y}}{\partial z} + \gamma_1 \frac{\partial Q_{1y}}{\partial x}\right], \tag{2.15}$$

where $k_{1xz}^2 = \gamma_1^2 - k_y^2$.

Similarly, from (2.4), with implicit $\exp(ik_y y)$ dependence on the variable $y$, we may obtain

$$Q_{2x} = \frac{1}{k_{2xz}^2}\left[ik_y \frac{\partial Q_{2y}}{\partial x} + \gamma_2 \frac{\partial Q_{2y}}{\partial z}\right], \tag{2.16}$$

$$Q_{2z} = \frac{1}{k_{2xz}^2}\left[ik_y \frac{\partial Q_{2y}}{\partial z} - \gamma_2 \frac{\partial Q_{2y}}{\partial x}\right], \tag{2.17}$$

with $k_{2xz}^2 = \gamma_2^2 - k_y^2$.

Equations (2.14), (2.15), (2.16) and (2.17) show that if the scalar fields $Q_{1y}$ and $Q_{2y}$ are evaluated then the vector fields $\underline{Q}_1$ and $\underline{Q}_2$ will be determined completely. Bearing this in mind, first comparing the $y$-component on both sides of (2.3), to get

$$\gamma_1 Q_{1y} = \frac{\partial Q_{1x}}{\partial z} - \frac{\partial Q_{1z}}{\partial x},$$

and then eliminating $Q_{1x}$ and $Q_{1z}$ from the above equation by using (2.14) and (2.15) we can obtain the following partial differential equation

$$\frac{\P^2 Q_{1y}}{\P x^2} + \frac{\P^2 Q_{1y}}{\P z^2} + k_{1xz}^2 Q_{1y} = 0. \qquad (2.18)$$

(2.18) shows that $Q_{1y}$ satisfies scalar Helmholtz equation.

Similarly we can derive

$$\left(\frac{\P^2}{\P z^2} + \frac{\P^2}{\P x^2}\right) Q_{2y} + k_{2xz}^2 Q_{2y} = 0. \qquad (2.19)$$

Since on a perfectly conducting surface, electric field vanishes, therefore the boundary conditions on a perfectly conducting half-plane in terms of the electric field components take the form $E_x = E_y = 0,$ for $z = 0, x \geq 0.$

By using these conditions in the *y*-components of $\underline{Q}_1$ and $\underline{Q}_2$ that is (2.5) and (2.6) we can obtain

$$Q_{1y} - ih_2 Q_{2y} = 0, \text{ and } Q_{1x} - ih_2 Q_{2x} = 0.$$

Further use of these two equations in (2.14) and (2.16) helps us to rewrite the boundary conditions in term of $Q_{1y}$ as

$$\frac{\P Q_{1y}}{\P x} \mp d \frac{\P Q_{1y}}{\P z} = 0, \quad z = 0^{\pm}, x \geq 0, \qquad (2.20)$$

where $d = \dfrac{g_2 k_{1xz}^2 + g_1 k_{2xz}^2}{ik_y\left(k_{2xz}^2 - k_{1xz}^2\right)}.$

*The reader might find it interesting to know that the boundary conditions (2.20) are of the same form as impedance boundary conditions as in Ref. [16].*

For $x<0$, $z=0$ there is no boundary therefore the continuity conditions are given by

$$Q_{1y}(x,z^{+}) = Q_{1y}\,(x,z^{-})\,;\;\; x<0, z=0, \tag{2.21}$$

$$\frac{¶Q_{1y}}{¶z}(x,z^{+}) = \frac{¶Q_{1y}}{¶z}(x,z^{-})\,;\;\; x<0, z=0. \tag{2.22}$$

Also a suitable edge condition is

$\lim_{r\to 0} Q_{1y} = C_1 + O(r)^{\frac{1}{2}}, \quad r=\sqrt{x^2+z^2},$ where $C_1$ is a constant.

The scattered field must satisfy the radiation conditions in the limit $r \to \infty$. We calculate only one field $Q_{1y}$ or $Q_{2y}$ because the presence of other scalar field can be seen from $Q_{1y} - ih_2 Q_{2y} = 0.$

Since the resultant vector field is equal to the sum of incident and scattered vector fields therefore

$$\underline{Q}_1 = \underline{Q}_1^{inc} + \underline{Q}_1^{sca}, \tag{2.23}$$

From (2.23) we can write

$$Q_{1y} = Q_{1y}^{\;inc} + Q_{1y}^{\;sca}. \tag{2.24}$$

The solution of (2.18) (*in a source-less field*) is discussed in [1].

## 3. Line source diffraction (asymptotically) by a perfectly conducting half-plane.

Let us extend the analysis in [1] to a line source at $(-x_0,-y_0)$ *asymptotically*. Therefore (2.18) will be replaced by

$$\left(\frac{\partial^2}{\partial x^2}+\frac{\partial^2}{\partial z^2}\right)\Theta_{1y}+k_{1xz}^2\Theta_{1y}=\delta(x+x_0)\delta(z+z_0), \tag{3.1}$$

where $k_{1xz}^2=\gamma_1^2-k_y^2$, and Dirac Delta function has the following property:

$$\delta(x+x_0)=0 \quad x\neq -x_0,$$

$$\neq 0 \quad x=-x_0.$$

From (2.24) and (3.1) it can be seen that the incident wave satisfies the following non-homogeneous scalar Helmholtz equation

$$\left(\frac{\partial^2}{\partial x^2}+\frac{\partial^2}{\partial z^2}\right)\Theta_{1y}^{inc}+k_{1xz}^2\Theta_{1y}^{inc}=\delta(x+x_0)\delta(z+z_0), \tag{3.2}$$

while the scattered wave $\Theta_{1y}^{sca}$ satisfies the following scalar homogeneous Helmholtz equation

$$\left(\frac{\partial^2}{\partial x^2}+\frac{\partial^2}{\partial z^2}\right)\Theta_{1y}^{sca}+k_{1xz}^2\Theta_{1y}^{sca}=0. \tag{3.3}$$

Defining the Fourier transform for the incident wave field $\Theta_{1y}^{inc}$ with respect to the variable $x$ as

$$\psi^{inc}(\vartheta,z)=\frac{1}{\sqrt{2\pi}}\int_{-\infty}^{+\infty}\Theta_{1y}^{inc}(x,z)e^{i\vartheta x}dx \tag{3.4}$$

$$=\psi_+^{inc}(\vartheta,z)+\psi_-^{inc}(\vartheta,z), \tag{3.5}$$

where the half-range Fourier transforms are given by

$$y_{+}^{inc}(J,z)=\frac{1}{\sqrt{2p}}\int_{0}^{\infty}Q_{1y}^{inc}(x,z)e^{iJx}dx, \tag{3.6}$$

$$y_{-}^{inc}(J,z)=\frac{1}{\sqrt{2p}}\int_{-\infty}^{0}Q_{1y}^{inc}(x,z)e^{iJx}dx. \tag{3.7}$$

The inverse Fourier transform is defined as

$$Q_{1y}^{inc}(x,z)=\frac{1}{\sqrt{2p}}\int_{-\infty}^{+\infty}y^{inc}(J,z)e^{-iJx}dJ. \tag{3.8}$$

On transforming (3.1), we obtain

$$\frac{d^2y^{inc}(J,z)}{dz^2}+\left(k_{1xz}^2-J^2\right)y^{inc}(J,z)=\frac{e^{-iJx_0}}{\sqrt{2p}}d(z+z_0), \tag{3.9}$$

Substituting $k^2=k_{1xz}^2-J^2$, we can write (3.9) as

$$\frac{d^2y^{inc}(J,z)}{dz^2}+k^2y^{inc}(J,z)=\frac{e^{-iJx_0}}{\sqrt{2p}}d(z+z_0), \tag{3.10}$$

where the right hand side of (3.10) is obtained by using the property of Dirac Delta function given by,

$$\int_{-\infty}^{+\infty}d(x+x_0)f(x)dx=f(-x_0).$$

The solution of the equation (3.10) has been discussed in Ref. [16] and is given by

$$y^{inc}(J,z)=\frac{\left(-\frac{1}{2}a\right)}{ik}\int_{-\infty}^{+\infty}e^{-ik|z-h'|}d(h'+z_0)dh', \tag{3.11}$$

where

$$a = \frac{1}{\sqrt{2p}} e^{-iJx_0}, \text{ and } \; k = \sqrt{k_{1xz}^2 - J^2},$$

so that

$$y^{\;inc}(J, z) = -\frac{e^{-iJx_0}}{2ik\sqrt{2p}} e^{-ik|z+z_0|}. \tag{3.12}$$

By taking the inverse Fourier transform of (3.12) we have

$$Q_{1y}^{inc}(x, z) = \frac{1}{\sqrt{2p}} \int_{-\infty}^{+\infty} y^{\;inc}(J, z) e^{-iJx} dJ.$$

After using (3.12) in the above equation we can obtain

$$Q_{1y}^{inc}(x, z) = -\frac{1}{4ip} \int_{-\infty}^{+\infty} \frac{e^{-iJ(x+x_0)-ik|z+z_0|} dJ}{k},$$

but $k = \sqrt{k_{1xz}^2 - J^2}$, therefore we have

$$Q_{1y}^{inc}(x, z) = -\frac{1}{4pi} \int_{-\infty}^{+\infty} \frac{e^{-iJ(x+x_0)-i\sqrt{k_{1xz}^2 - J^2}|z+z_0|} dJ}{\sqrt{k_{1xz}^2 - J^2}}. \tag{3.13}$$

By substituting

$J = k_{1xz} \cos(q + it)$, where $0 < q < p$, $-\infty < t < +\infty$, $|z + z_0| = R \sin q$,

$x + x_0 = R \cos q$, and

$$R = \sqrt{(x + x_0)^2 + (z + z_0)^2}, \tag{3.14}$$

the right hand side of (3.13) will take the form

$$Q_{1y}^{inc}(x,z) = \frac{1}{4p}\int_{-\infty}^{+\infty} e^{-ik_{1xz}R\cosh t}\,dt$$

$$\Rightarrow Q_{1y}^{inc}(x,z) = -\frac{i}{4}H_0^{(2)}(k_{1xz}R), \tag{3.15}$$

which is the Hankel function (second kind) of order zero.

By using the *asymptotic* value of $H_0^{(2)}(k_{1xz}R)$, (3.15) can be written as

$$Q_{1y}^{inc}(x,z) = -\frac{i}{4}\sqrt{\frac{2}{pk_{1xz}R}}\,e^{-i\left(k_{1xz}R-\frac{p}{4}\right)}, \tag{3.16}$$

where *R* is given by (3.14).

Now substituting $-x_0 = r_0\cos f_0,\ -z_0 = r_0\sin f_0$ in (3.14), the approximate value of *R* for a very large value of $r_0$ is given by

$$R \approx r_0 - (x\cos f_0 + z\sin f_0), \text{ where } -p < f_0 < -\frac{p}{2}. \tag{3.17}$$

Therefore by using (3.17) the incident wave (3.16) can be written as

$$Q_{1y}^{inc}(x,z) \approx -\frac{i}{4}\sqrt{\frac{2}{pk_{1xz}r_0}}\,e^{-i\left(k_{1xz}r_0-\frac{p}{4}\right)}e^{ik_{1xz}(x\cos f_0 + z\sin f_0)}. \tag{3.18}$$

By writing

$$-\frac{i}{4}\sqrt{\frac{2}{pk_{1xz}r_0}}\,e^{-i\left(k_{1xz}r_0-\frac{p}{4}\right)} = c, \tag{3.19}$$

(3.18) can be reduced to

$$Q_{1y}^{inc}(x,z)=ce^{ik_{1xz}(x\cos f_0+z\sin f_0)}, \qquad \text{or} \tag{3.20}$$

$$Q_{1y}^{inc}(x,z)=ce^{i(xk_{1x}+zk_{1z})}, \tag{3.21}$$

where $k_{1x}=k_{1xz}\cos f_0$ and $k_{1z}=k_{1xz}\sin f_0$ (*asymptotically*) and the constant $c$ is given by (3.19). In (3.21) the $y$ dependence has been suppressed. (3.21) shows that the incident wave behaves like a plane wave after applying the far field approximation.

*Now the problem under discussion is quite similar to the one in [1].* As we proceed further the reader is advised to see [1] for the missing mathematical details.

## 4. The Wiener-Hopf Equations For $\boxed{Q_{1y}}$

In order to solve the equation (3.3) the required Wiener-Hopf equations are

$$y'_-(J,0)-ikL(J)\bar{j}_+(J,0)+\frac{ck_{1z}}{\sqrt{2p}\,(J+k_{1x})}=0, \tag{4.1}$$

and

$$-iJy_-(J,0)+dL(J)\bar{j}'_+(J,0)+\frac{c}{\sqrt{2p}}\frac{k_{1x}}{(J+k_{1x})}=0, \tag{4.2}$$

where $L(J)=1+\dfrac{J}{kd}$, (4.3)

$$\bar{j}_+(J,0)=\frac{1}{2}\left[y_+(J,0^+)-y_+(J,0^-)\right],$$

and $\bar{j}'_+(J,0)=\dfrac{1}{2}\left[y'_+(J,0^+)-y'_+(J,0^-)\right].$

The prime denotes the differentiation with respect to the variable *z*.

The two Wiener-Hopf equations (4.1) and (4.2) are to be solved for the unknown functions $\overline{\dot{J}}_+(J,0)$ and $\overline{\dot{J}'_+}(J,0)$ by the Wiener-Hopf technique.

*As a reminder the detailed derivations of* (4.1) *and* (4.2) *could be found in* [1].

## 4.1 Solution of the Wiener-Hopf Equations for $Q_{1y}$

The function $L(J)$ given by equation (4.3) can be factorized as

$$L(J) = L_+(J)L_-(J), \tag{4.4}$$

where $L_+(J)$ is regular in the upper half plane and $L_-(J)$ is regular in the lower half-plane.

Equation (4.4) is taken as a definition but the explicit forms of $L_\pm(J)$ are available in the Appendix of [1].

As in [1] on adopting the usual Wiener-Hopf procedure for (4.1) and (4.2) and using the extended form of Liouville's theorem [16] the unknowns are given by

$$\overline{\dot{J}}_+(J,0) = \frac{c}{\sqrt{2p}} \frac{k_{1z}}{i(J + k_{1x})\sqrt{k_{1xz} + k_{1x}}\,L_-(-k_{1x})\sqrt{k_{1xz} + J}\,L_+(J)} \tag{4.5}$$

and

$$\overline{\dot{J}'_+}(J,0) = -\frac{c}{\sqrt{2p}}\left[\frac{k_{1x}}{dL_+(J)L_-(-k_{1x})(J + k_{1x})}\right]. \tag{4.6}$$

Finally, the diffracted field can be written as [1]

$$Q_{1y}^{sca}(x,z) = \frac{1}{\sqrt{2p}} \int_{-\infty}^{+\infty} G(J) e^{ikz} e^{-iJx} dJ\,, \tag{4.7}$$

where

$$G(J) = \frac{c}{\sqrt{2p}}\left[\frac{k_{1z}}{i(J + k_{1x})\sqrt{k_{1xz} + k_{1x}}\,L_{-}(-k_{1x})\sqrt{k_{1xz} + J}\,L_{+}(J)}\right]$$
$$-\frac{c}{ik\sqrt{2p}}\left[\frac{k_{1x}}{dL_{+}(J)L_{-}(-k_{1x})(J + k_{1x})}\right]. \qquad (4.8)$$

## 4.2 $Q_{1y}^{sca}(x,z)$ in the far zone

The contribution to the right side of equation (4.7) from the simple pole of the function $G(J)$ at $J = -k_{1x}$ can be calculated by closing the contour of integration by a semicircle in the lower half plane. If we denote this contribution by $Q_{1y}^{sca\prime}(x,z)$ we obtain

$$Q_{1y}^{sca\prime}(x,z) = -c\exp i(k_{1x}x + k_{1z}z), \qquad (4.9)$$

where $c$ is given by the equation (3.19). (4.9) is negative of the incident wave field (3.21).

To determine the far zone behavior of the diffracted field, we introduce the polar coordinates (r, θ) via

$x = r\cos q$ , $y = r\sin q$ .

Using the transformation

$J = -k_{1xz}\cos(q + it)$, where $t$ is real, the integral in equation (4.7) can be evaluated asymptotically by the saddle point method.

Thus, omitting the details of calculation, the far zone field is given by

$$Q_{1y}^{sca}(x,z) \sim G(-k_{1xz}\cos q)\left[\frac{p}{2k_{1xz}r}\right]^{\frac{1}{2}}\exp i\left(k_{1xz}r - \frac{p}{4}\right), \qquad (4.10)$$

where $G(-k_{1xz}\cos\theta)$ can be found from (4.8). Finally $Q_{1y}(x,z)$ can be obtained from (2.24) after using (3.21) and (4.10). Similarly the right-handed Beltrami field $Q_{2y}(x,z)$ can be determined from

$$\frac{\partial^2 Q_{2y}^{inc}}{\partial x^2}+\frac{\partial^2 Q_{2y}^{inc}}{\partial z^2}+k_{2xz}^2 Q_{2y}^{inc}=\delta(x+x_0)\delta(z+z_0),$$

and $$\frac{\partial^2 Q_{2y}^{sca}}{\partial x^2}+\frac{\partial^2 Q_{2y}^{sca}}{\partial z^2}+k_{2xz}^2 Q_{2y}^{sca}=0.$$

## 5. Conclusions

(1) As in [1] the vector diffraction problem is reduced to the scattering of a single scalar field (being the normal component of either a left handed or right handed Beltrami field) and is contrary to the assertion in Ref. [17].

(2) Another important observation is that the problem of diffraction of electromagnetic waves by a perfectly conducting half-plane is similar to that of scattering of electromagnetic waves by an imperfectly conducting half-plane (satisfying impedance boundary conditions) in an otherwise homogeneous medium.

Related material could be found in References [16] and [19]. Thus, this observation leads to the conclusion that the whole class of these problems in electromagnetic and acoustic theory can be tackled for the bi-isotropic medium.

(3) [1] Opens a new area of diffraction in bi-isotropic medium, it is advisable to extend this field further. This paper is an attempt in that direction, in which diffraction of line source electromagnetic wave by a perfectly conducting half-plane in a bi-isotropic medium has been considered asymptotically.

(4) In addition, applications, related with the propagation of plane waves with **negative phase velocity** (i.e. *velocity opposite to the direction of power flow*) in isotropic chiral materials are also very interesting to look at. Related work could be found in References [11], [12], [13] and [14].

(5) The observations presented in this paper will be a gateway to address many more problems of great practical importance for example *point source diffraction.* Scattering by a perfectly conducting obstacle in a homogeneous chiral environment could be found in Ref. [2]. See also References [3] and [15] for more interesting results.